\documentclass[11pt]{article}
\usepackage[margin=1in]{geometry}
\usepackage{cite}
\usepackage{graphicx}
\usepackage{xcolor}
\usepackage{wrapfig}

\begin{document}

\begin{center}
{\LARGE Profit from Two Bitcoin Mining Tactics:\\ Towing and Shutdown}
\end{center}

\begin{center}
Ehsan Meamari ~~~~~~ Chien-Chung Shen\\
Department of Computer and Information Sciences \\
University of Delaware, U.S.A.\\
\{ehsan,cshen\}@udel.edu
\end{center}

\section{Introduction}

Since Bitcoin's inception in 2008, it has became attractive investments for both trading and mining. To mine Bitcoins, a miner has to invest in computing power and pay for electricity to solve cryptographic puzzles for rewards, if it becomes the first to solve a puzzle, paid in Bitcoin. Given that mining is such a resource intensive effort, miners seek new strategies trying to make the mining process more profitable. One obvious strategy is to adopt faster and/or more energy-efficient compute hardware \cite{Taylor-Evolution-of-Bitcoin-Hardware}. In addition, miners could launch attacks, such as Selfish \cite{Sapirshtein-Optimal-Selfish-Mining-Strategies-in-Bitcoin} and Withholding \cite{Eyal-Miners-Dilemma}, to earn more Bitcoins. In this article, we introduce two new tactics termed Shutdown and Towing and analyze their profitability of earning more Bitcoins. In the following, we first review a simple background, and then present the two tactics.

\section{Review of key Bitcoin concepts}

There is no central authority in the Bitcoin system to approve transactions. Therefore, miners collect valid transactions in a (local) block, add a Coinbase transaction for a mining award, compete to solve a puzzle, and then, upon solving a puzzle, broadcast the valid block to the network. Later, all other miners will accept the new valid block and append it to the existing blockchain.~As the probability of individual miners' winning the puzzle-solving competition diminishes when more miners are competing, mining pools are formed to mine cooperatively. To articulate Towing and Shutdown tactics, we consider a simplified Bitcoin network of five mining pools.

As shown in \cite{hash-rate}, the total hash-rate ($H_{total}$) ({\em hashes/second}) of the Bitcoin network fluctuates over time. Puzzles could be solved sooner when $H_{total}$ increases, and vice versa. Such fluctuation may vary the average block generation time which is expected to be around 10 minutes, according to the Bitcoin protocol. To address this issue, Bitcoin adjusts the mining difficulty ($D$) of solving puzzles according to $H_{total}$ so as to maintain the 10-minute average block generation time.

\section{Towing tactic}

\begin{figure*}[h]
\centering
\includegraphics[width=1\textwidth]{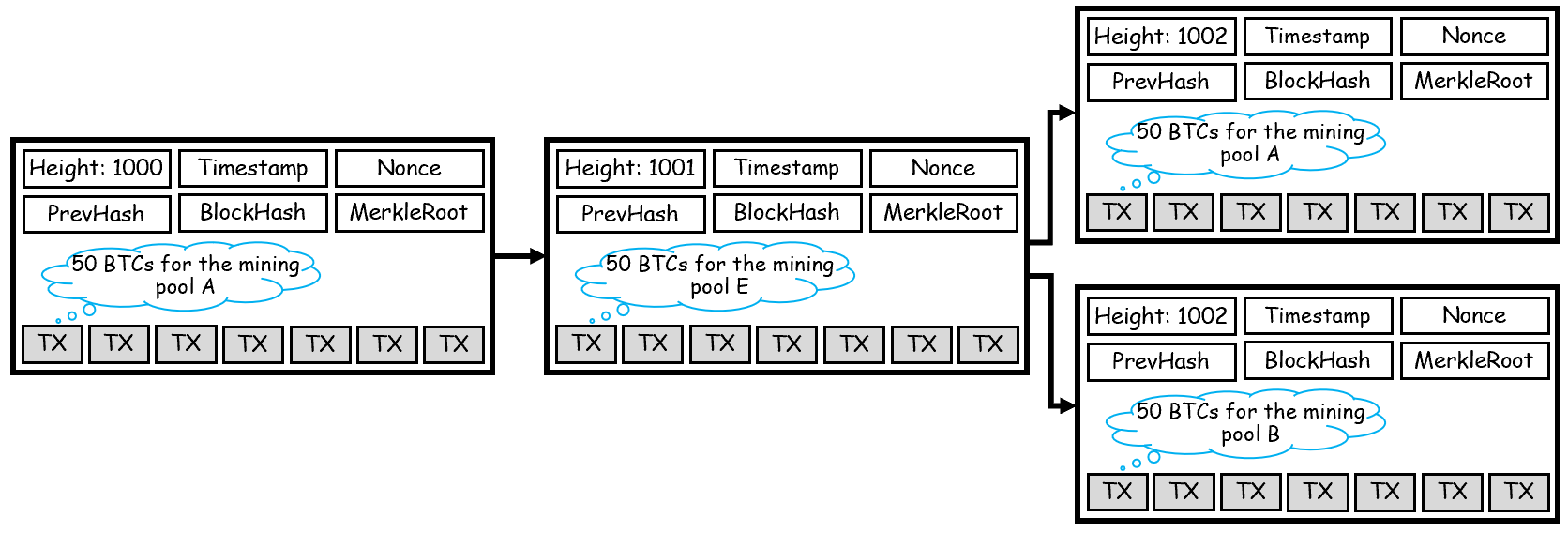}
\caption{A Split Scenario}
\label{fig:split}
\end{figure*}

This section introduces the idea of Towing tactic. Let $R$ denotes the reward of each block generation, and $C$ the average cost for solving a puzzle. Then the net utility of each block mining is $R-C$. Suppose that the block (with height) 1000 had been mined by the mining pool $A$, as depicted in Fig.~\ref{fig:split}.
Therefore, in its Coinbase transaction, the mining pool $A$ received the reward of 50 BTCs (We do not consider transaction fee, for simplicity.) and it earns net utility 40 BTCs if we consider $C=10$ BTCs.

Fig.~\ref{fig:split} shows that the block 1001 was mined by the mining pool $E$. After a miner solves a puzzle, it needs to disseminate the valid block to the Bitcoin network. For various reasons, such as network latency, some of the miners/pools do not receive a newly mined block and continue mining, and hence a blockchain split could happen such as the two blocks of the same height 1002 in Fig.~\ref{fig:split}, where the block in the upper branch is mined by the mining pool $A$ and the block in the lower branch by the mining pool $B$.

The issue now is that which branch should be agreed upon as the new blockchain head by the entire network. In the Bitcoin protocol, the method to resolve this dilemma is that miners accept the first received block and reject the later ones with the same height. Therefore, assuming the block generated by the mining pool $A$ is propagated sooner, more miners probably will accept it and try to continue mining on it. Sometimes, both branches continued to be the head for more mining and additional blocks were added to them. However, eventually one branch would be longer and miners accept the longer branch so that the shorter branch is discarded to become orphan blocks. Although this split may persist into block 1003, we assume that the shorter branch becomes orphan after (the first) block 1003 has been added into the blockchain. Let $H_{i}$ be the sum of hash-rates of the miners/mining pools who try to append a block 1003 onto branch $i$. The probability of branch $i$ winning this competition ($P_{i}$) is $H_{i}/H_{total}$.

Suppose that mining pools $B$ and $C$ each has 20\% of $H_{total}$ and three other mining pools have the remaining 60\% of $H_{total}$. Without loss of generality, let's say that mining pool $B$ will continue with the lower branch to mine block 1003, then $P_{lower}=20\%$. Other mining pools choose the upper branch, since they received block 1002 from the mining pool $A$ sooner. Therefore, the mining pool $B$ expect to get $P_{lower} \cdot (R-C) =8~(BTCs)$ for block 1003.

If a mining pool wants to increase its expected net utility, it could increase the chance of winning by attracting more miners to its branch. However, the Bitcoin protocol advises miners/pools to work on the first received block. Therefore, when a miner receives a message from a peer proposing a new block with the same height, it will reject this new block. Instead, mining pools can apply Towing tactic with the prerequisite that two mining pools had agreed that one pool will help the other when the latter sticks in a split. Therefore, if a mining pool is mining on a block and receives a new message proposing a block with the same height from its accomplice, it will give up the current mining activity to accept the new block and mine on it.

In our simple network, suppose that mining pools $B$ and $C$ have agreed to apply the Towing tactic. On the split in Fig. \ref{fig:split}, the mining pool $C$ sees the opportunity to apply the Towing tactic and support the mining pool $B$. This means that although the mining pool $C$ received the block of the upper branch earlier, it gives up the upper branch and cooperate with the mining pool $B$ to mine on the lower branch and double the new $P_{lower}$. Therefore, the expected utility for the mining pool $B$ will be increased to 16 (BTCs).

\section{Temporary Shutdown tactic}

This section discusses the profitability of using temporary Shutdown tactic to earn more Bitcoin. We term each sequence of 2,016 blocks as a {\em mining period} ($P$) so that the very first 2,016 blocks on the Bitcoin blockchain would be $P^1$, the subsequent 2,016 blocks on the Bitcoin blockchain would be $P^2$, and so on. Let $t^j$ be the time (in days) taken to generate all the 2,016 blocks of mining period $P^j$, which is supposed to be two weeks, as mining each block is expected to take 10 minutes on average.

When $H_{total}$ of the Bitcoin network increases, new blocks could be generated sooner than 10 minutes, so that $t^j$ could be less than two weeks. Therefore, before starting the subsequent mining period, mining difficulty should be adjusted to a higher value to decelerate the block generation in the next mining period, and vice versa. In the Bitcoin protocol, the mining difficulty for mining period $P^j$ ($D^j$) changes at the end of mining period $P^{j-1}$, and is inversely proportional to $t^{j-1}$, i.e., $D^j=D^{j-1}\cdot(14~days)/t^{j-1}$. We denote $H_{total}^j$ to be the average total hash-rate of the Bitcoin network during mining period $P^j$ and $H_i^j$ the average total hash-rate of mining pool $i$ during mining period $P^j$. The {\em daily expected utility} for mining pool $i$ in mining period $P^j$ ($DEU_i^j$) is calculated by Eq. \ref{DEU}, where $r_i^j$ is the ratio of $H_i^j$ over $H_{total}^j$.

\begin{equation}
    DEU_i^j=(r_i^j\cdot(R-C)\cdot2016)/t^j
    \label{DEU}
\end{equation}

\begin{figure}[h]
\centering
\includegraphics[width=0.5\textwidth]{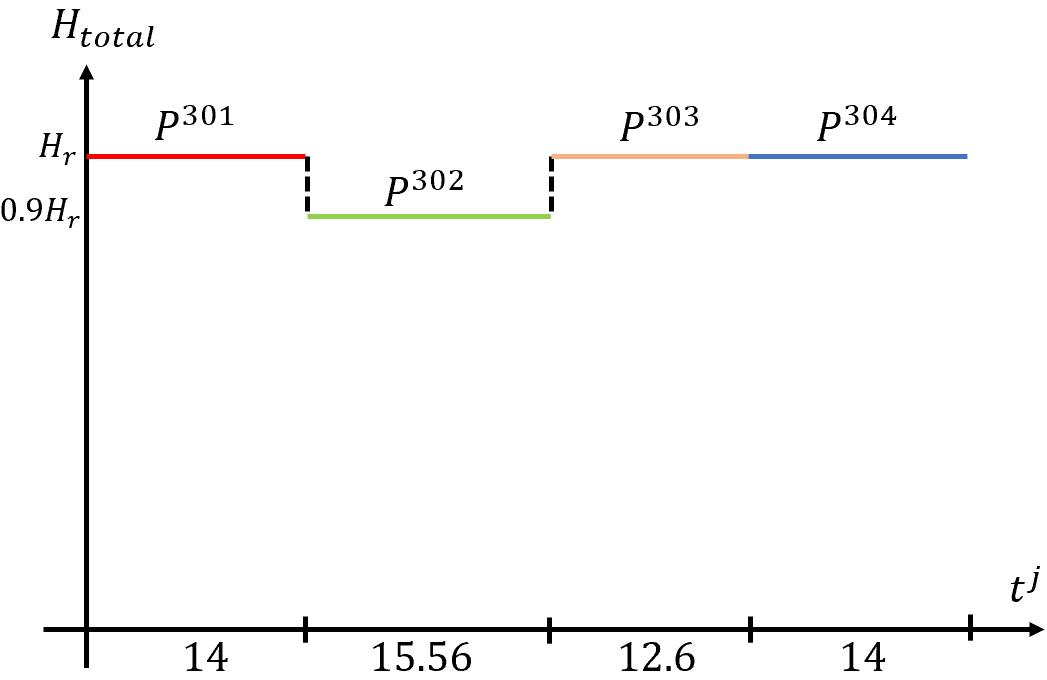}
\caption{Temporary Shutdown in a simple scenario,}
\label{fig:period}
\end{figure}

To demonstrate the potential profitability of the Shutdown tactic, we analyze a simple scenario as depicted in Fig. \ref{fig:period}. We assume that $R$ is 12.5 BTCs for mining period $P^{300}$, and do not consider transaction fees. As a reference, we assume that it take all the mining pools, with a {\em reference total hash-rate} $H_r$, a {\em reference mining cost} $C_r$ of 11.5 BTCs, and a {\em reference mining difficulty} $D_r$, to complete mining period $P^{301}$ in 14 days.
We supposed before that there are five mining pools forming our simplified Bitcoin network, and $r_i^{301}=20\%$ for each mining pool $i$. Therefore, $DEU_i^{301}$ computes to 28.80 BTCs, according to Eq. \ref{DEU}, for each mining pool $i$, as shown across the `None' row in Table \ref{table:ComputationResults}.

As mining difficulty $D^{j+1}$ is adjusted (indirectly) according to $H_{total}^{j}$, by applying the Shutdown tactic, a mining pool shuts down a portion of its compute power to decrease the $H_{total}$ of the Bitcoin network, which increases the block generation time, and hence, simplifies the puzzles for the subsequent mining period. As the result, the net profit increases as solving simpler puzzles costs less. 

For instance, suppose that, upon starting $P^{302}$, the mining pool $A$ shuts down half of its compute power to decrease its corresponding hash-rates to $H_A^{302}=0.1H_r$, as depicted in Fig.~\ref{fig:period}, so that $r_A^{302}=(0.1H_r)/(0.9H_r)$. As $t^{301}$ is 14 days, the mining difficulty $D^{302}$ stays the same from the previous period. Lower $H_{total}^{302}$ (= 0.9$H_r$) prolongs the mining of new blocks in $P^{302}$ to 11.11 minutes each, so that $P^{302}$ takes 15.56 days to mine its 2,016 blocks. At the end of $P^{302}$, as $t^{302}$ is longer than two weeks, the Bitcoin protocol decreases its mining difficulty for $P^{303}$ to $D^{303}=0.9D_r$,
which decreases the cost to $0.9C_r$ or 10.35 BTCs. Then, upon starting the mining period $P^{303}$, mining pool $A$ turns the shutdown compute power back on to restore $H_A^{303}$ back to 0.2$H_r$, which speeds up puzzle solving and decreases $t^{303}$ to $12.6$ days.

Row `ST' in Table \ref{table:ComputationResults} shows that when mining pool $A$ shuts down half of its compute power, $DEU$ (averaged over both mining periods of $P^{302}$ and $P^{303}$) for all the mining pools increases. The Improvement Values (IV) denote that $DEU$ of mining pool $A$ increases by 35\%, and 62\% for all the other mining pools. This shows that although other mining pools did not shut down any portion of their compute power, they gain more profits than mining pool $A$. As $D^{304}=1.1D^{303}=D_r$, the mining cost is $C_r$ and $DEU$ computes to 28.80 BTCs during the mining period 304, similar to the mining period 301.

\begin{table}[t]
\begin{tabular}{| *{16}{c|} }
\hline
& \multicolumn{3}{c|}{Mining Pool A}
& \multicolumn{3}{c|}{Mining Pool B}
& \multicolumn{3}{c|}{Mining Pool C}
& \multicolumn{3}{c|}{Mining Pool D}
& \multicolumn{3}{c|}{Mining Pool E}   \\
\hline
Mode   &   H  &   DEU  &   IV &   H  &   DEU  &   IV &   H  &   DEU  &   IV &   H  &   DEU  &   IV &   H  &   DEU  &   IV  \\
\hline\hline
None   &  20   &  28.80   &  0   &  20   &      28.80   &  0    &  20   &      28.80   &  0   &  20   &  28.80   &  0    &  20   &   28.80   &  0         \\
\hline
ST   &  10   &  38.74   &  35   &  20   &  46.7   &  62   &  20   &     46.7   &  62    &  20   &    46.7   &  62    &  20   &     46.7   &  62     \\
\hline
\end{tabular}
\caption{Results for using the shutdown tactic}
\label{table:ComputationResults}
\end{table}

\section{Future Work}

In this article, we introduced the Towing and Shutdown tactics and demonstrated how mining pools could increase their net mining utilities by applying these two tactics. For a comprehensive analysis of the Shutdown tactic, we need to answer more complicated questions. For instance, with the execution of the Shutdown tactic by mining pools, would new independent miners be incentivized to join the Bitcoin network and start mining, which might nullify the tactic? In addition, could the scale of shutdown be optimized to maximize mining profits? To analyze the Towing tactic in the actual Bitcoin network, we would need to consider more accurate reward and cost values, and the role of delay in disseminating blocks. 

\bibliographystyle{IEEEtran}
\bibliography{Reference}

\begin{wrapfigure}{l}{25mm} 
\includegraphics[width=1in,height=1.25in,clip,keepaspectratio]{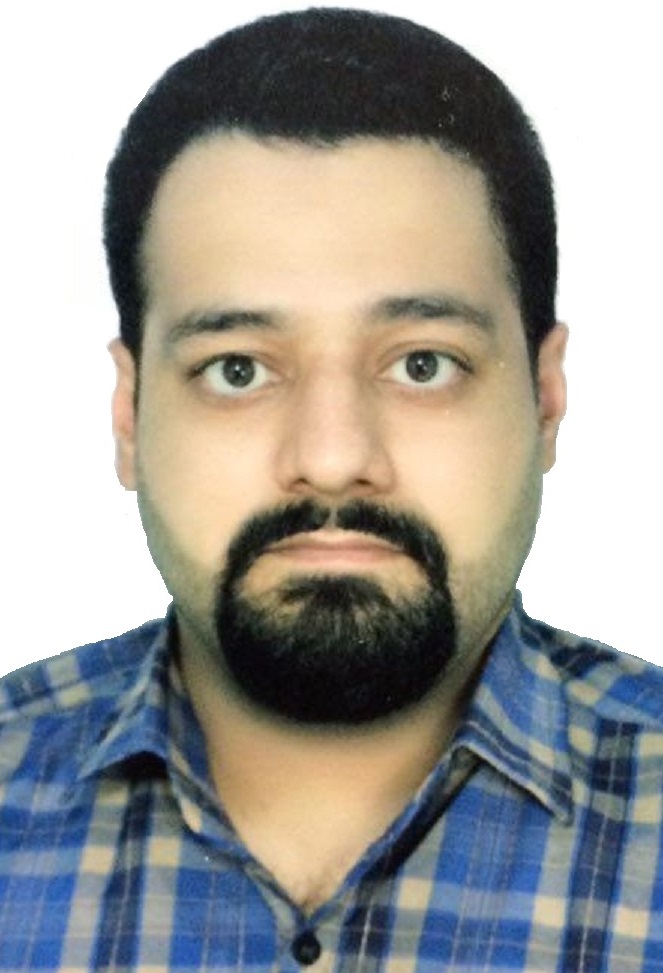}
\end{wrapfigure}\par
\textcolor{white}{//}

\textbf{Ehsan Meamari} received his B.Sc. degree in Telecommunication Engineering from the University of Birjand in 2009 and an M.Sc. degree in Information Technology Engineering from Iran University of Science and Technology in 2012. He is currently a Ph.D. student in the Department of Computer and Information Sciences at the University of Delaware. His research interests include (but not limited to) network security, cryptography, artificial intelligence, and Blockchain.\par
\textcolor{white}{//}

\textcolor{white}{//}

\begin{wrapfigure}{l}{25mm} 
\includegraphics[width=1in,height=1.25in,clip,keepaspectratio]{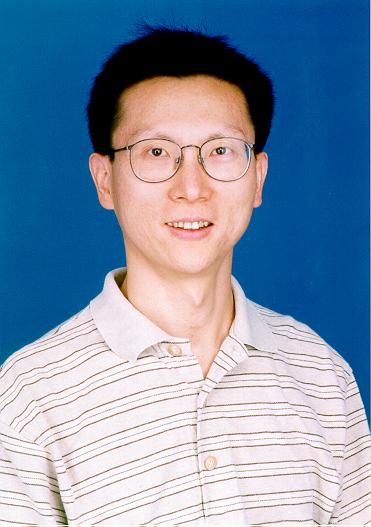}
\end{wrapfigure}\par
\textbf{Chien-Chung Shen} received his B.S. and M.S. degrees from National Chiao Tung University, Taiwan, and his Ph.D. degree from UCLA, all in computer science.  He was a research scientist at Bellcore Applied Research working on control and management of broadband networks.  He is now a Professor in the Department of Computer and Information Sciences of the University of Delaware. His research interests include blockchain, Wi-Fi, SDN, NFV, ad hoc and sensor networks, dynamic spectrum management, cybersecurity, and simulation.  He is a recipient of NSF CAREER Award and a member of both ACM and IEEE.

\end{document}